\documentstyle[11pt,paspconf,psfig]{article}
\def\edcomment#1{\iffalse\marginpar{\raggedright\sl#1\/}\else\relax\fi}
\reversemarginpar
\newcommand{\ltapprox}{\raisebox{-0.5ex}{$\,\stackrel{<}{\scriptstyle\sim}\,$}}
\newcommand{\gtapprox}{\raisebox{-0.5ex}{$\,\stackrel{>}{\scriptstyle\sim}\,$}}

\begin{document}
\title{Photometric Redshifts in lensing clusters: Identification and Calibration
of faint high-$z$ galaxies}
\author{Roser Pell\'o and Jean-Paul Kneib}
\affil{Observatoire Midi-Pyr\'en\'ees, UMR 5572, 14, Avenue E. Belin, F-31400 
Toulouse, France}
\author{Micol Bolzonella}
\affil{Istituto di Fisica Cosmica "G.P.S. Occhialini", via Bassini 15, 20133 Milano, Italy}
\author{Joan-Marc Miralles}
\affil{Astronomical Institute, Tohoku University, Aramaki, Aoba-ku, Sendai 980-8578, 
Japan}

\begin{abstract}

We discuss a method aiming to use photometric redshifts in lensing clusters to
access the population of distant background sources. The amplification provided
by gravitational lensing allows to calibrate photometric redshifts 1 to 3 magnitudes
deeper with respect to field studies. We present and discuss some results obtained 
with this procedure on extensively studied clusters, as well as the future issues 
of this project.

\end{abstract}


\section{Introduction}

Lensing clusters can be used as gravitational telescopes to build up and to
study an independent sample of high-$z$ galaxies, in order to complement the
large samples obtained in field surveys. The amplification close to the critical lines
is typically $\Delta m \sim $ 2 to 3 mags, and it is still $\sim 1$ mag at 1' 
from the cluster center (see Figure~1). The signal/noise ratio 
in spectra of amplified sources
and the detection fluxes are improved beyond the limits of conventional techniques, 
whatever the wavelenght used for this exercise. 
In particular, the amplification properties have been succesfully used
in the ultra-deep MIR survey of A2390 (Altieri et al. 1999), and the SCUBA
cluster lens-survey (Smail et al 98; Blain et al. 99).
We discuss here the
systematic use of photometric redshifts in lensing clusters to identify 
high-$z$ sources. This is the basis of a large collaboration program 
presently going on, and involving different european institutions, aimimg to 
perform the spectroscopic follow up of high-$z$ candidates selected from the 
visible and near-IR photometry.

The first lensed galaxy confirmed at $z \gtapprox 2$ was the spectacular
blue arc in Cl2244-02 (Mellier et al., 1991). More recent
examples of highly-magnified galaxies, identified either purposely or
serendipitously, strongly encourages this approach: the star-forming source
$\#384$ in A2218 at z=2.51 (Ebbels et al. 1996); the luminous z=2.7 arc 
behind the EMSS cluster MS1512+36 (Yee et al. 1996; Seitz et al., 1998); three z
$\sim$ 4 galaxies in Cl0939+47 (Trager et al. 1997); a z=4.92 system in
Cl1358+62 (Franx et al. 1997, Soifer et al. 1998); and the two red galaxies at
$z \sim 4$ in A2390 (Frye \& Broadhurst 1998, Pell\'o et al. 1999).

\begin{figure}
\psfig{figure=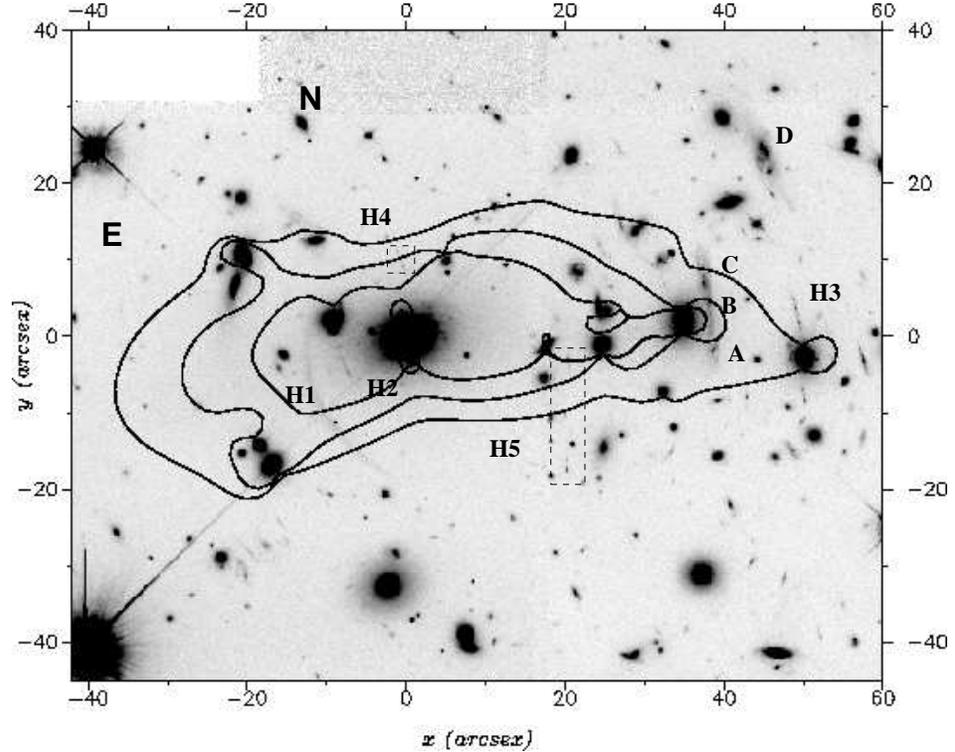,angle=270,height=10.0cm}
\label{fig1}
\caption{WFPC2 image of A2390 (F814W) showing the location of the 
critical lines at $z$ $=$ 1.0, 2.5 and 4. Several multiple and 
high redshift images are identified}
\end{figure}

\section{Photometric Redshifts}

The method used here to compute photometric redshifts (hereafter {\it $z_{phot}$}) is a
standard SED fitting procedure, according to:

\begin{equation}
 \chi^2(z)=\sum_{i=1}^{N_{filters}} \left({F_i [Observed]
- b \times F_i [Template](z)\over \sigma_i} \right) ^2
\label{eq1}
\end{equation}

where $ F_i [Observed]$, $F_i [Template]$, and  $\sigma_i$ are the 
observed and template fluxes and their uncertainty in filter i, respectively, and
b is a normalisation constant. It was originally developed by Miralles (1998)
(see also Miralles \& Pell\'o 1998), and a new version of this code called 
{\it hyperz} is presently developed (Bolzonella et al. in preparation).
The set of templates includes mainly spectra from the Bruzual \& Charlot evolutionary code
(GISSEL98, Bruzual \& Charlot, 1993), and also a set of empirical SEDs compiled by
Coleman, Wu and Weedman (1980) to represent the local population of galaxies. 
The synthetic database derived from Bruzual \& Charlot includes 255 spectra, distributed into
5 different star-formation regimes (51 different
ages for the stellar population, all of them with solar metallicity): a burst of 0.1
Gyr, a constant star-formation rate, and 3 $\mu$ models (exponential-decaying SFR) with 
characteristic times of star-formation chosen to match the present-day sequence of E, 
Sa and Sc galaxies. The reddening law is taken from Calzetti (1999), but 4 other laws are 
also included in our code. The normal setting for $A_v$ ranges between 0 and 0.5 magnitudes. 
Flux decrements in the Lyman forest are computed 
according to Giallongo \& Cristiani (1990) or Madau (1995), both of them giving similar
results. When applying {\it hyperz} to the spectroscopic samples available on the HDF,
the uncertainties are typically $\delta z / (1 + z) \sim 0.1$.

\section{Identification of high-$z$ sources}

\begin{figure}
\psfig{figure=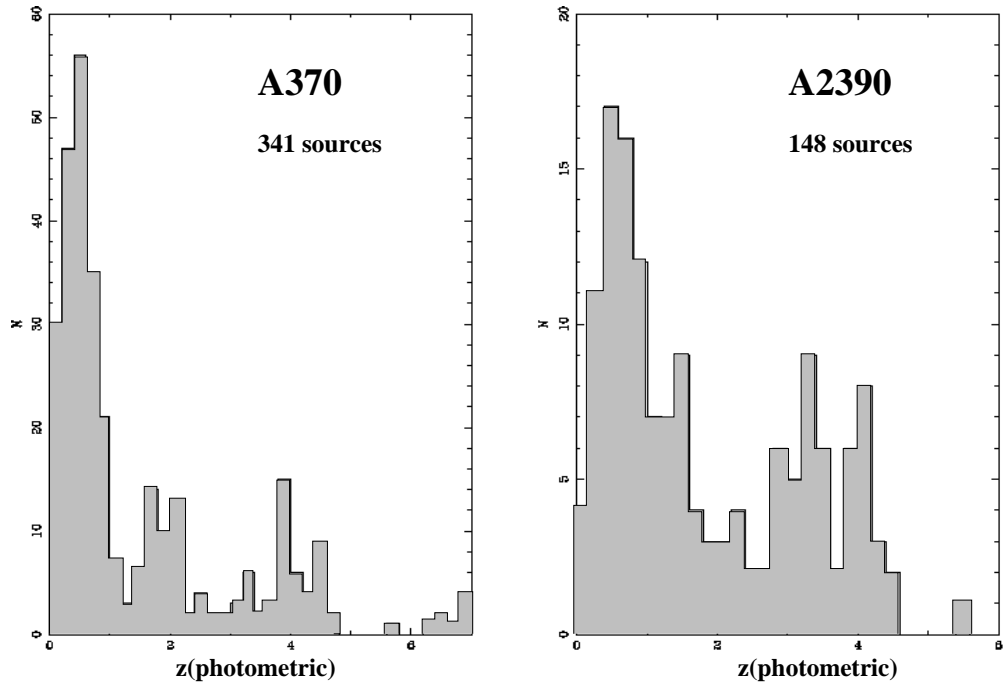,angle=270,height=9.0cm}
\label{fig2}
\caption{{\it $z_{phot}$} distribution of arclets in A2390 (right) and 
A370 (left), obtained with {\it hyperz}. In A2390, selection criteria  
are based on the morphology of the WFPC2 images (elongation, orientation 
and surface brightness); in A370, only a photometric selection was applied,
aimed to avoid the obvious bright cluster members.}
\end{figure}

High-z lensed sources are selected close to the appropriate critical lines
(see Figure~1), with {\it $z_{phot}$} $\ge 2$. Figure~2 displays
the {\it $z_{phot}$} distribution of arclets in two well known lensing clusters,
where the sample of arclets have been applied according to different selection 
criteria. In all cases, {\it $z_{phot}$} is computed on the basis of a 
photometric survey including near-IR data. The method is restricted to 
lensing clusters whose mass distribution is highly constrained by multiple 
images (revealed by HST or ground-based multicolor images), where the amplification
uncertainties are typically $\Delta m_{lensing}<$ 0.3 magnitudes.
Such clusters with well constrained mass distributions enable to
recover precisely the properties of lensed galaxies (morphology,
magnification factor). Highly magnified sources are presently the only 
way to access the dynamical
properties of galaxies at $z \ge 2$, through 2D spectroscopy, at a spatial resolution
$\sim$ 1 kpc. The two multiple-images at the same $z \sim 4$, observed behind
A2390, are un example of these reconstruction capabilities (Pell\'o et al. 1999).
Cluster lenses can be used advantageously to determine the redshift
distribution up to the faintest levels through magnified sources
(Figure~2).
It is also the natural way to search for primeval galaxies, in order to
constraint the scenarios of galaxy formation.

\begin{figure}
\vbox{
\psfig{figure=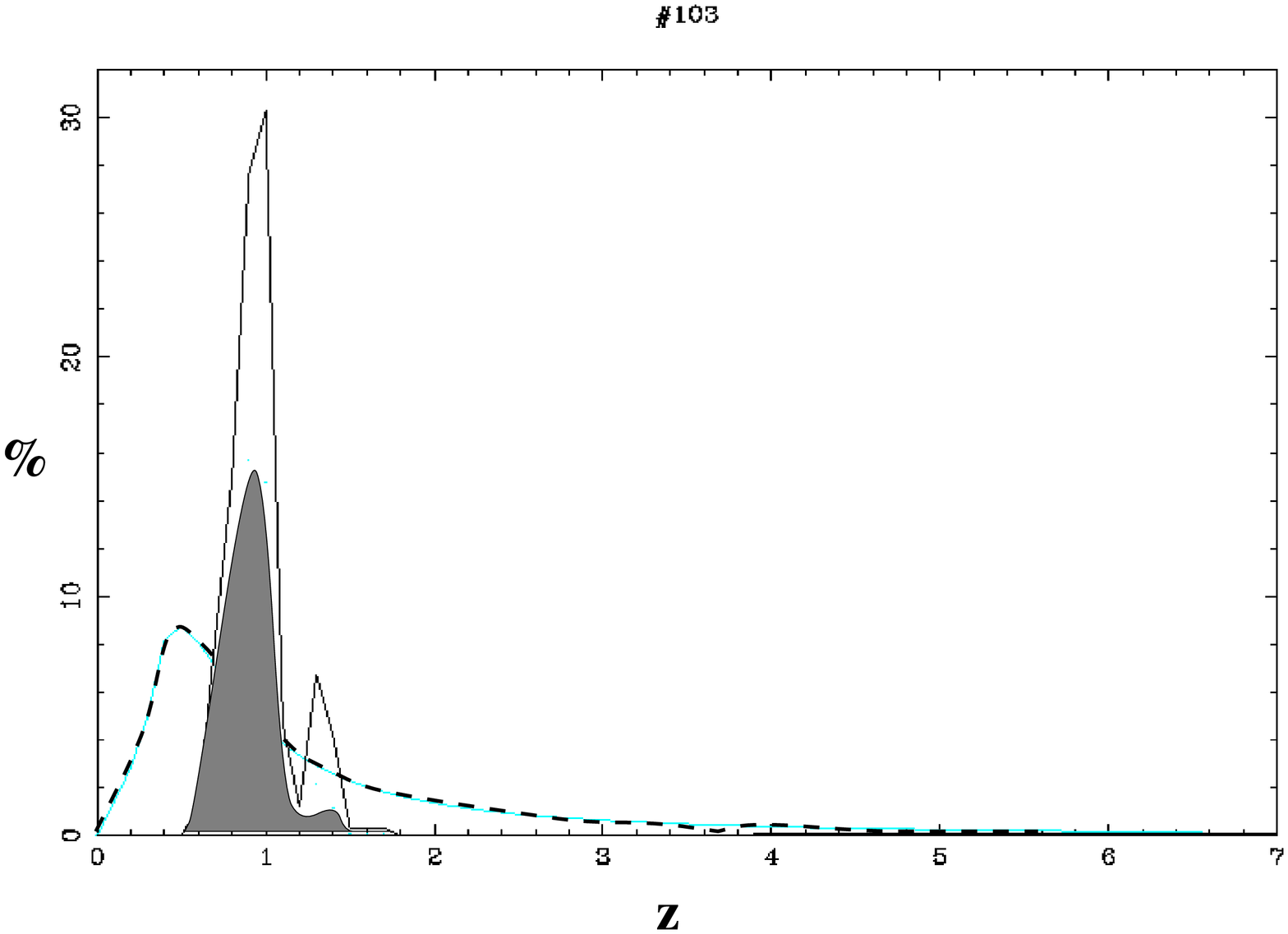,angle=270,height=6.0cm}
\psfig{figure=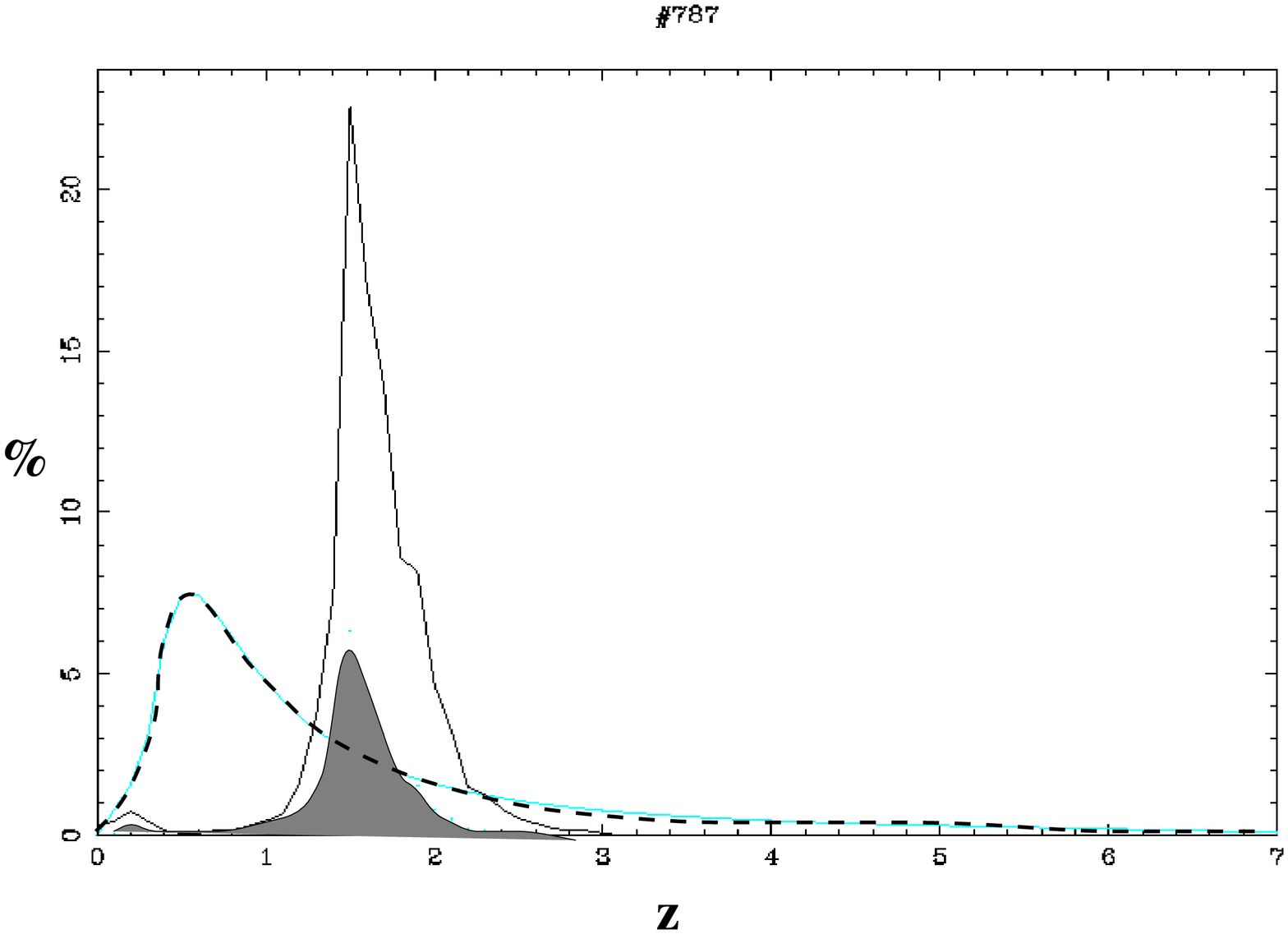,angle=270,height=6.0cm}
}
\label{fig3}
\caption{Examples of redshift probability distributions for 2 ISO sources in
A2390, \# 103 (left) and \# 787 (right). Dashed lines give the lens inversion likelihood 
functions, whereas solid lines correspond to the {\it $z_{phot}$} probability distributions.
Shaded regions correspond to the final composite probability.
}
\end{figure}

\section{Photometric Redshifts compared to Spectroscopic and Lens Inversion results}

For a subsample of spectroscopically confirmed objects, we have tested the
photometric redshift accuracy as a function of the relevant parameters (SFR,
reddening, age and metallicity of the stellar population). We have also cross-checked
the consistency between the photometric, the spectroscopic and the lensing redshift
obtained from inversion methods (Ebbels et al. 1998). The agreement
between the three methods is good up to at least $z \ltapprox 1.5$. For
higher redshifts, the results for the most amplified sources are
promising, but an enlarged spectroscopic sample is urgently needed. The comparison
between the spectroscopic and the lensing redshift has been studied in the field
of A2218 (Ebbels et al. 1998), and all the present results seem to follow this trend 
at least to $z \ltapprox 1.5$.
Figure~4 displays the difference
between {\it $z_{phot}$} and lens redshift for a subsample of 98 arclets in the core of 
A2390, selected according to morphological criteria (minimum elongation and 
right orientation are requested). According to these results, about $60\%$ of the sample
have $| \Delta z | \le 0.25 $. Most of the discrepancy corresponds to sources
with $z \gtapprox 2$. 
A general trend exists with high-$z$ images, which are not
correctly identified by inversion techniques as compared to {\it $z_{phot}$}.
This behaviour is expected as a result of the relative low sensitivity to $z$ of the 
inversion method for high-$z$ values.

Taking into account that {\it $z_{phot}$} and lensing inversion techniques produce 
independent probability distributions for amplified sources, combining both methods 
provides with an alternative way to determine the redshift distribution
of high-$z$ sources. Figure~3 shows un example for 2 MIR selected sources in A2390. 

\begin{figure}
\psfig{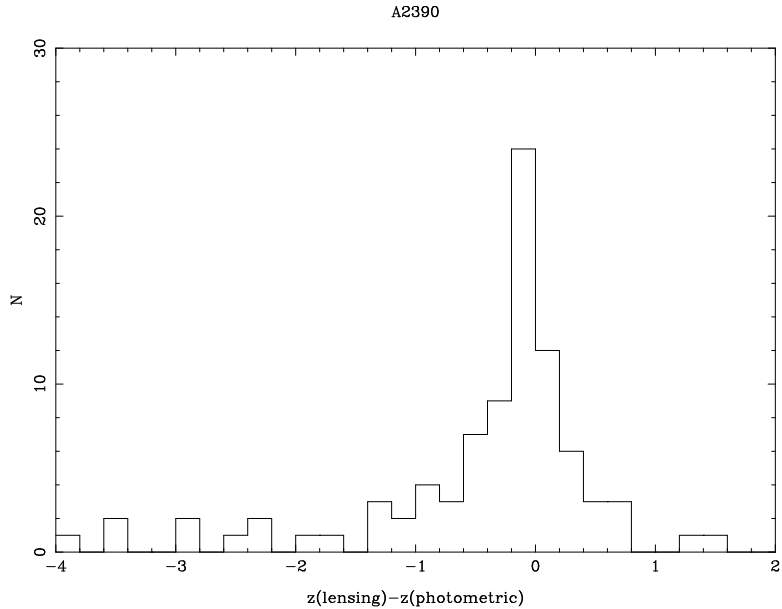}
\label{fig5}
\caption{Difference between {\it $z_{phot}$} and lens redshift for a sample of 98
arclets in the core of A2390 (morphological selection). }
\end{figure}

\section{Conclusions and Future Developments}

The selection of high-$z$ candidates in cluster lenses using a {\it $z_{phot}$} approach
is strongly supported by the present results. The efficiency is
increased when using a large wavelength range to compute {\it $z_{phot}$},
including the IR bands. 
For most statistical purposes, {\it $z_{phot}$}  should be accurate enough to
discuss the properties of these extremely distant galaxies. Conversely, 
lensing clusters could be used as a tool to check photometric redshifts up to the 
faintest limits, through the spectroscopic confirmation of {\it $z_{phot}$}
for such amplified sources. An Ultra-Deep Photometric Survey of selected cluster lenses is 
urgently needed to probe the distant Universe, and this could be a 
a well defined program for the ACS camera on HST.

\acknowledgments

We are grateful to G. Bruzual, S. Charlot, Y. Mellier, B. Fort, R.S. Ellis, 
J.F. Le Borgne and M. Dantel-Fort for useful discussions on this particular technique.
Part of this work was supported by the TMR {\it Lensnet} ERBFMRXCT97-0172
(http://www.ast.cam.ac.uk/IoA/lensnet).

\end{document}